# Predicting Variation of DNA Shape Preferences in Protein-DNA Interaction in Cancer Cells with a New Biophysical Model

Kirill Batmanov and Junbai Wang *

Department of Pathology, Oslo University Hospital—Norwegian Radium Hospital, Montebello, 0310 Oslo, Norway; Kirill.Batmanov@rr-research.no
* Correspondence: junbai.wang@rr-research.no

**Abstract:** DNA shape readout is an important mechanism of transcription factor target site recognition, in addition to the sequence readout. Several machine learning-based models of transcription factor–DNA interactions, considering DNA shape features, have been developed in recent years. Here, we present a new biophysical model of protein–DNA interactions by integrating the DNA shape properties. It is based on the neighbor dinucleotide dependency model BayesPI2, where new parameters are restricted to a subspace spanned by the dinucleotide form of DNA shape features. This allows a biophysical interpretation of the new parameters as a position-dependent preference towards specific DNA shape features. Using the new model, we explore the variation of DNA shape preferences in several transcription factors across various cancer cell lines and cellular conditions. The results reveal that there are DNA shape variations at FOXA1 (Forkhead Box Protein A1) binding sites in steroid-treated MCF7 cells. The new biophysical model is useful for elucidating the finer details of transcription factor–DNA interaction, as well as for predicting cancer mutation effects in the future.

## 1. Introduction

Understanding how transcription factors (TFs) recognize their target DNA binding sites is an important task in the study of gene regulation. Although a complete model of this process is currently out of reach [1], the growing body of experimental data enables the development of many approximate models to compute TF–DNA binding affinity. There are models that aim at identifying proteins that may bind to DNA based on the protein amino acid sequences (e.g., nDNA-Prot [2]) or models that focus on the prediction of protein target sites (e.g., BayesPI2 [3]). The latter ones are useful in identifying functional TF binding sites, predicting the effects of mutations on gene regulation [4], and elucidating the differences between related TFs [5]. Current approaches to estimate the TF–DNA binding affinity can be broadly divided into two categories: one is the "black box" approach, which uses powerful machine learning techniques (e.g., support vector machines, random forest, neural network or ensemble methods such as LibD3C [6]) with as many input features as possible, to achieve the most accurate affinity predictions [7–10]; and the other is the biophysical modeling approach, which derives physical models from first principles by using well-understood approximations in statistical physics [11–13]. Generally, machine learning methods have high accuracy after considering many diverse input features, but they do not consider the domain knowledge. Thus, it is difficult to interpret the results that are learned from training data. Additionally, experimental biases [14,15] may be unintentionally learned by the model parameters, which makes a reliable test prediction only happen for experiments with the same configuration as that of training ones. In this work, we pursue the second approach, biophysical modeling, which produces interpretable model parameters relevant to the TF–DNA interaction mechanisms. The trained model parameters have clear definitions in theory. Due to such advantages, the new biophysical model can be adjusted and reused in different settings, and its parameter values can be compared between models. Many of the previous biophysical models of TF–DNA interactions do not consider the contribution of dinucleotide dependency in TF binding sites [13,16]. The inferred TF binding target motif is parameterized by a position-specific weight matrix (PWM), which is simple and interpretable. The independent TF binding model is often used to investigate biological phenomena in gene regulation. For example, a PWM-based biophysical model, BayesPI2, has been applied to study subtle transcription factor binding patterns and gene regulation effects in cancer cells [17]. Another similar program, BayesPI-BAR, has been used to predict the significance of mutation effects on TF–DNA binding, which led to the discovery of new regulatory mutations that cause gene dysregulation in follicular lymphoma [18].

In addition to the nucleotide sequence, local DNA structure properties, such as the geometry of the DNA molecule and base stacking energy, are known to affect the protein–DNA interaction [19–21]. Using such properties as input features together with the DNA sequence has improved TF–DNA binding prediction accuracy in several machine learning-based models [9,22–24]. However, until now, there is no biophysical model incorporating DNA structure information. Several works have considered general dinucleotide dependency features [3,11,12], which implicitly contain DNA shape information [22]. The results have been modest, with some evidence showing that the dinucleotide dependency models do not generalize well to in vivo data [25]. Here, we have developed a new biophysical model that considers DNA structure features as a special form of position-specific dinucleotide dependency. This model has its parameters restricted to a space defined by the dinucleotide combination of DNA structure properties derived from the DiProDB database [26]. We compare the performance of the independent PWM model, the full dinucleotide model and the DNA shape-restricted dinucleotide model on several tasks such as single nucleotide variant (SNV) effect analysis, which requires accurate TF binding affinity prediction. Unlike the machine learning approaches, the new biophysical model equipped with interpretable model parameters and the training structure, the inferred shape feature preferences can be further studied in various conditions. It gives us an opportunity to investigate the dynamical change of DNA shape preferences in different cancer cell lines.

## 2. Materials and Methods

### 2.1. ChIP-Seq Data

The ChIP-seq data used to fit the TF–DNA binding affinity models was downloaded from the Encyclopedia of DNA Elements (ENCODE) project [27]. We use peaks from the uniform peak calling pipeline. The ER$\alpha$ (Estrogen Receptor Alpha) time series ChIP-seq is obtained from Gene Expression Omnibus (GEO), Accession GSE94023. FOXA1 ChIP-seq for different conditions is obtained from [28] (GEO Accession GSE72249). We take up to 1000 peaks with the highest signal value from each experiment, and for each peak, a 100 bp sequence centered on the called peak is extracted from the hg19 reference genome. These sequences form the positive set of samples, which are assumed to frequently contain the binding motif for the corresponding TF. To fit an affinity model, our methods also need a set of negative samples. This set consists of 100 bp sequences taken randomly from genome regions near transcription start sites (TSS) of known genes (TSS ± 10 kbp). The peak regions are excluded from the negative set, and the number of sequences in each negative set is the same as the size of the positive set.

### 2.2. Allele-Specific Binding Data

The dataset of allele-specific TF binding events was downloaded from a recent publication [29]. It was derived from raw reads of ENCODE ChIP-seq experiments for 36 TFs in several cell lines, where SNVs were first identified and then assessed for possible allele-specific binding based on the number of reads containing each variant.

### 2.3. BayesPI Transcription Factor–DNA Binding Affinity Model

The basic biophysical model of TF–DNA binding affinity, called BayesPI, was first presented in [13] and later extended in [3] to include the contribution of interactions between neighboring nucleotides in the DNA. The expression for the probability of TF binding to a small DNA segment is:

$$P(S, w, \mu) = \sum_{i=0}^{N-M} \frac{1}{1 + e^{E(S_{i:i+M}, w) - \mu}}$$

where $S_{i,a} = 1$ if the DNA sequence has nucleotide $a$ (one of A, C, G, T) at position $i$ and $S_{i,j} = 0$ otherwise, $N$ is the sequence length, $M$ is the length of the binding motif (the number of consecutive base pairs that affect the affinity), and $\mu$ is the chemical potential of the TF, which is defined by its concentration in the nucleus. The binding energy of the TF to a short DNA fragment with length $M$ bp is represented by $E$, which is the sum of the independent contributions of each nucleotide:

$$E_{indep}(S, w) = \sum_{j=0}^{M-1} \sum_{a=1}^{4} w_{j,a} S_{j,a}$$

The matrix $w \in \mathbb{R}^{M \times 4}$, called the position-specific affinity matrix, specifies these contributions: $w_{j,a}$ is the binding energy of nucleotide $a$ at position $j$ inside the DNA fragment. The protein binding energy may consider the contribution of dinucleotide dependence:

$$E_{dinuc}(S, d) = \sum_{j=0}^{M-2} \sum_{a=1}^{4} \sum_{b=1}^{4} d_{j,(a-1)*4+b} S_{j,a} S_{j+1,b}$$

where $d \in \mathbb{R}^{M-1 \times 16}$ is the matrix of pairwise dependency energy correction, with $d_{j,(a-1)*4+b}$ specifying the correction of the independent energy terms for the dinucleotide at positions $j: j+1$ with nucleotides $a$ and $b$.

### 2.4. DNA Shape Affinity Model

The TF–DNA binding affinity model that takes into account DNA shape information includes dinucleotide dependencies in a more structured form, following prior information about DNA molecule characteristics, which may be important for the protein–DNA interaction. The affinity is modeled as:

$$P(S, F, w, d^f, \mu) = \sum_{i=0}^{N-M} \frac{1}{1 + e^{E_{indep}(S_{i:i+M}, w) + E_{shape}(F_{i:i+M}, d^f) - \mu}}$$

where $F \in \mathbb{R}^{M \times K}$ is the matrix of DNA shape feature values at each position inside the sequence $S$ and $K$ is the number of different DNA shape features considered. $F_{i,j}$ is the value of DNA shape feature $j$ at position $i$, which is a number characterizing a certain property of the DNA molecule at that location. Analogous to the independent and dinucleotide models, DNA shape contributes linearly to the binding energy:

$$E_{shape}(F, d^f) = \sum_{j=0}^{M-1} \sum_{k=1}^{K} d_{j,k}^f F_{j,k}$$

where $d^f \in \mathbb{R}^{M \times K}$ is the matrix of position-specific DNA shape preference of the TF, expressed as a correction to the independent nucleotide preferences. The 2–mer shape features used in this work come from the DiProDB database [26], which lists thermodynamic, structural and some other properties of DNA and RNA dinucleotides. We use four features: twist, minor groove width, propeller twist and roll. These features are chosen following an earlier DNA shape publication [30], where the same DNA shape characteristics are used in TF–DNA affinity modeling. However, the DNA shape features in [30] are based on 5– and 6–mer sequences, while DiProDB uses dinucleotides only. In the dinucleotide case, $F$ can be expressed as:

$$F_{j,k} = \sum_{a=1}^{4} \sum_{b=1}^{4} D_{k,(a-1)*4+b} S_{j,a} S_{j+1,b}$$

where $D \in \mathbb{R}^{K \times 16}$ is the matrix of $K$ DNA shape feature values for each dinucleotide. In our model, $K = 4$. Plugging the expression for $F$ into the definition of $E_{shape}$, we observe that, as it is defined here, the DNA shape energy contribution $E_{shape}$ and the dinucleotide dependency energy contribution $E_{dinuc}$ are linearly related:

$$E_{shape}(F, d^f) = E_{dinuc}(S, d^f \cdot D)$$

In other words, the DNA shape model is a dinucleotide dependency model whose coefficients are restricted to a linear subspace spanned by the shape features $D$. This is a useful property that allows efficient computation of affinity without the need to compute DNA shape features explicitly for each sequence. Furthermore, any software that works with dinucleotide models, such as BayesPI2, can now be used to work with DNA shape models by expanding $d^f$ coefficients into full dinucleotide coefficients $d = d^f \cdot D$.

*2.5. Bayesian Inference of Model Parameters for the DNA Shape Restricted Dinucleotide Dependence Model*

The parameters of TF–DNA affinity models ($w$, $\mu$ and $d^f$) are fitted to the experimental data, such as ChIP-seq peaks or protein binding microarray probes, using gradient descent, starting from a randomized initial seed. Because the experimental data typically contain measurement noise, a proper regularization of the model is important [31]. This is done by applying Bayesian inference to the L2 regularization hyperparameters, hence the name BayesPI. The Bayesian inference proceeds in iterations of sequential estimation of model parameters, such as $w$, $\mu$ and $d^f$, and hyperparameters, which are L2 regularization coefficients $\alpha_i$ and the error scaling coefficient $\beta$. When fitting the model parameters, the gradient descent is performed on the following regularized cost function:

$$L = \beta L_D(w, d^f, \mu, a, b) + \sum_i \alpha_i W_i^2$$

$$L_D(w, d^f, \mu, a, b) = \frac{1}{2} \sum_i \left((a \cdot P(S_i, w, \mu, d^f) + b) - t_i\right)^2$$

where $t_i$ is the target value (e.g., normalized ChIP-seq tag count) for sequence $S_i$, $a$ and $b$ are coefficients of the linear transformation applied to the binding probability in order to match its distribution to the distribution of $t_i$ and $W$ is the sequence of all model parameters: $w$, $d^f$, $\mu$, $a$ and $b$. On the first iteration, the regularization hyperparameters $\beta$ and $\alpha_i$ are set to constant values corresponding to a weak regularization. After the gradient descent has converged, the hyperparameter values are updated using the following formulas:

$$\alpha_i = \frac{\gamma_i}{W_i^2}$$

$$\beta = \frac{N - \sum_i \gamma_i}{2 L_D}$$

$$\gamma_i = 1 - \alpha_i H_{ii}^{-1}$$

where $N$ is the number of sequences and $H = J(\nabla L)^T$ is the Hessian matrix of the cost function, evaluated at the converged values of model parameters, and $J$ is the Jacobian matrix (see the Supplementary Methods for the formulas to compute the Hessian). Since $\gamma_i$ and $H$ depend on $\beta$ and $\alpha_i$, the hyperparameter update is also performed iteratively, repeatedly recomputing values of $\gamma_i$ and $H$, followed by recomputation of $\beta$ and $\alpha_i$, until convergence. The values of model parameters $W$ are kept fixed during the hyperparameter update computation.

After the hyperparameters have been updated, we again perform the gradient descent on the model parameters, starting from their values on the previous iteration. The whole process is repeated several times, usually converging in 3–10 iterations. The derivation of this algorithm, based on Bayesian treatment of the uncertainty of the hyperparameters, is given in [32]. In BayesPI, the $\alpha_i$ hyperparameters are grouped together and their values shared between classes. All weights of the independent part of the model, $w$, share a single $\alpha_w$ (please refer to the Supplementary Methods for computational details of this procedure). The dinucleotide DNA shape feature weights at the same position $p$, $d_{p,k}^f$ also share the same regularization weight $\alpha_{d,p}$. The reason is that the information content of the motif is distributed unequally across positions, with some positions being important while others affecting the affinity weakly. To avoid the overfitting of larger shape model parameters, they should be more strongly regularized at unimportant positions. This is achieved automatically by the Bayesian regression procedure. Since all dinucleotide weights at the same position share a single hyperparameter value, they will all be regularized with the same strength. Position-based regularization allows restricting the dependency parameters at positions where the dependency does not seem to contribute to the affinity, while letting the parameters grow at more important positions. The model is typically fit in two stages: first, the independent parameters $w$ are fit, setting $d^f = 0$. Then, $w$ is fixed, and the dependent parameters $d^f$ are fit. Thus, the dinucleotide shape-restricted dependency parameters $d^f$ should be viewed as adjustments of the independent model, in places where dependencies are significant. $d^f = 0$ recovers the original independent model.

*2.6. Mutation Effect Prediction by Using Various Biophysical Models*

We use the BayesPI-BAR [4] method to evaluate how a DNA variant affects TF binding. BayesPI-BAR computes a score called shifted differential binding affinity ($\delta dbA$) for each variant and TF. $\delta dbA$ measures the difference in above-background binding strength between the reference and alternate sequences, which is equivalent to measuring the effect of the sequence variant on TF binding. If $\delta dbA > 0$, then there is an increase of TF binding affinity in the alternate sequence compared to that in the reference sequence: creation of a new TF binding site or strengthening of an existing weak binding site. If $\delta dbA < 0$, then an existing TF binding site is disrupted by the variant. In the previous works, we only considered the independent binding affinity model for TFs in BayesPI-BAR. Here, we use independent, full dinucleotide dependence, and newly-developed DNA shape-restricted dinucleotide dependence models to evaluate the prediction of the mutation effect in cancer cells.

*2.7. Code Availability*

The new BayesPI2Shape software contains independent, dinucleotide dependence, and DNA shape-restricted dinucleotide dependence models (preselected DNA shape features and the visualization function) are available from http://folk.uio.no/junbaiw/bayesPI2shape/.

**3. Results**

*3.1. Validation of Inferred DNA Shape Feature Preferences for Protein–DNA Interaction*

The main advantage of the new shape-restricted TF–DNA affinity model is the interpretability of the model parameters. The matrix $d^f$ models the adjustments of the binding energy, due to the preferred DNA shape feature at a specific TF binding position: a positive value of $d^f_{j,k}$ indicates that the higher the values of shape feature $k$ at position $j$, the higher the TF binding affinity; but a negative value indicates that a low value of shape feature is preferred at the TF binding position. Thus, we call $d^f$ the (DNA) shape preference matrix. To validate the fitted shape preferences, we tested the new method on three TFs (Serum Response Factor (SRF), Myocyte Enhancer Factor 2C (MEF2C), and TATA-Box Binding Protein (TBP)) for which the DNA shape preferences were previously reported. In this test, we use PWMs from JASPAR and fit only $d^f$ using the ChIP-seq datasets from ENCODE. In Figure 1A, a heatmap of the shape feature preferences of SRF is shown. The most pronounced shape feature preference is for higher propeller twist at position 6, which matches the findings in [9]. The exact position may be shifted a few base pairs compared to Figure 7 in [9], because 5– and 6–mer DNA shape features are used in the previous work, while we use only dinucleotide shape features.

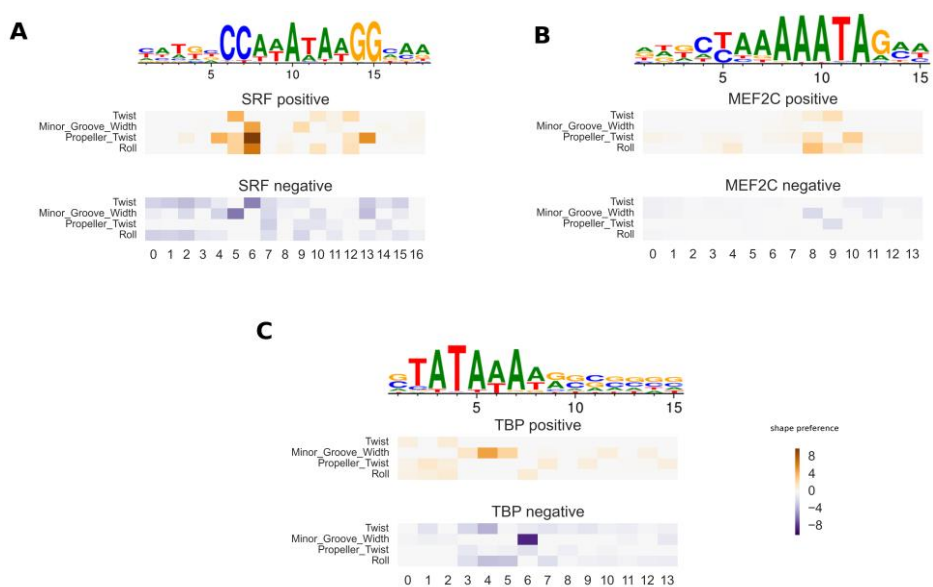

**Figure 1.** Shape feature preferences for Serum Response Factor (SRF), Myocyte Enhancer Factor 2C (MEF2C), and TATA-Box Binding Protein (TBP). The heatmaps show the preference for each shape feature at each binding position (the inferred $d^f$ matrix), with the preference for low and high feature values in blue and orange, respectively. The shape preferences are split into positive and negative parts accordingly. Grey color represents weaker preference. The position-specific weight matrix (PWM) logos of transcription factors (TFs) are shown above the shape preference heatmaps.

In Figure 1B, the heatmap for MEF2C is shown, where propeller twist and roll have strong preferences in the current prediction. The two shape features are also the most important features for MEF2C reported in [9], although their locations and strength are slightly different. It is noteworthy that in [9], the relative feature importance score given by the random forest model was used to predict DNA shape feature preferences. This score cannot be compared across different models. On the other hand, our biophysically-modeled DNA shape preferences $d^f$ directly reflect the dynamical modification of TF binding affinities in the change of DNA shape features. In Figure 1C, our predicted TBP shape feature preferences are shown, where minor groove width is the main dependency term. The result is consistent with the fact that TBP recognizes its target sequence by binding to the minor groove [33].

*3.2. ChIP-Seq Peak Prediction*

Here, we evaluate the performance of three affinity models (the independent model, the full dinucleotide model and the DNA shape-restricted dinucleotide model) to predict ChIP-seq peaks, using the ENCODE data. First, three independent models with motif sizes of 10, 15 and 20 were fitted for each of 36 TFs. We retain only the models that fit the training data well: $r^2$ of prediction for the raw signal is greater than 0.4, the area under receiver operating characteristic curve (AUC) >0.75 for distinguishing true peak sequences from the background ones. Then, we fit the dinucleotide interaction models and DNA shape models as corrections to the independent models on the same training data. The prediction results on an independent test set are shown in Figure 2. The AUC of each independent model is plotted against the AUC of the same model with a dependency correction. Both the dinucleotide dependent and the DNA shape-restricted models improve the accuracy of peak prediction significantly, which are compared to that by the independent model. In the current study, the full dinucleotide model (Figure 2A) usually has much higher accuracy than the DNA shape model (Figure 2B). The improvement of TF binding motif prediction by using the nucleotide dependent models (e.g., dinucleotides, *k*-mers and DNA shape features), on a homogeneous dataset such as ChIP-seq or protein binding microarray data, has been demonstrated before with black box machine learning techniques [9,12,25]. Here, we show that it also holds true for the new biophysical model. Figure 2 indicates that the full dinucleotide model is superior to the DNA shape-restricted dinucleotide model. However, such outcomes need to be considered with caution. That is because an experiment such as ChIP-seq is known to contain biases in the measurements. These biases do not reflect the underlying true biochemical processes, but may belong to a systematic error from the experimental procedure. Such errors may be learned by a powerful machine learning method. For example, the full dinucleotide model may learn a spurious dependency originating from an experiment-specific bias, which leads to the model only performing well in data generated from the same type of experiments. To fully assess the usefulness of a new model, it needs to be validated by different types of experimental datasets (e.g., a model trained on in vivo data, but tested on in vitro data).

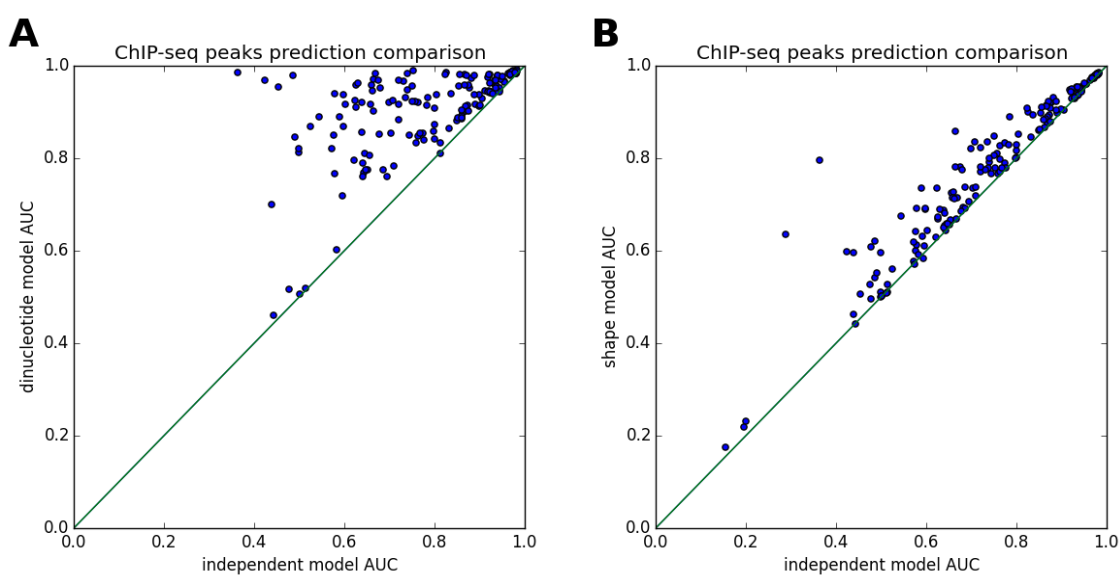

**Figure 2.** ChIP-seq peaks prediction is improved in dependency of the models. For each of the 172 ChIP-seq datasets for 36 TFs, the area under receiver operating characteristic curve (AUC) of the independent model is compared to (**A**) the full dinucleotide model, which has a separate tunable parameter for each of 16 possible dinucleotides at each position, and (**B**) the DNA shape-restricted dinucleotide model, which has a separate tunable parameter for each shape feature (e.g., four DNA shape features in the current study) at each position.

*3.3. Investigating the Variation of Predicted DNA Shape Preferences across Cell Types*

The parameters of the new DNA shape model have a straightforward interpretation, similar to that of the dinucleotide dependency model: each $d_{j,k}^f$ is the weight of the preference of the TF for the shape feature $k$ at position $j$, in addition to that defined by the independent model weights $w$. This enables an easy comparison of fitted shape parameters between different conditions. Here, we explore the variation of fitted shape model parameters, across different cancer cell lines. Our model is based on the hypothesis that the DNA shape readout by a TF depends only on the TF itself and the local DNA sequence. This assumption may be violated in vivo due to epigenetic DNA modifications [23,34], nucleosome positioning in the DNA sequence [35] or because of the presence of cofactors [20]. In order to assess the influence of these external factors on the shape model fitting procedure, we compared shape features predicted in different cell lines and different conditions for the same TF.

We use canonical PWMs from the JASPAR web database [36] for 23 TFs, where ENCODE ChIP-seq datasets are available for more than one cell line. For each TF, we fit the shape model parameters for each ChIP-seq dataset separately and compute the correlation of the fitted shape preference parameters between different cell lines. The distribution of the median correlation coefficients across TFs is shown in Figure 3A. Three of the TFs (Basic Helix-Loop-Helix Family Member E40 (BHLHE40,) MYC Associated Factor X (MAX), and JunD Proto-Oncogene (JUND)) have very low shape feature correlations between the conditions (e.g., <0.2). The PWM that we used for JUND has a poor performance for distinguishing the true peaks from the background ones: AUC is 0.58. This suggests that if the independent model cannot predict TF bindings well, then the result of the shape model is also unreliable. On average, the correlation of predicted shape preferences among different cell lines is high, with the median of them being around 0.7. The shape preference seems stable in *in vivo* conditions. In Figure 3B,C, the fitted shape feature variations in three conditions are given for E74 Like ETS Transcription Factor 1 (ELF1) (median correlation 0.7) and Upstream Transcription Factor 2 (USF2) (median correlation 0.67), respectively, where each shape model matrix $d^f$ is displayed by a heatmap (positive preferences). The negative preferences are shown in Supplementary Figures S2A,B. In Figure 4, MAX and BHLHE40 have nearly identical PWMs with very pronounced 6 bp core motifs, and the independent parts of the binding affinity models are very similar, though the inferred shape preferences for MAX and BHLHE40 differ significantly across cell lines (see Supplementary Figure S1 for the positive shape preferences). Thus, the predicted shape preferences may be used to distinguish the true binding sites between MAX and BHLHE40.

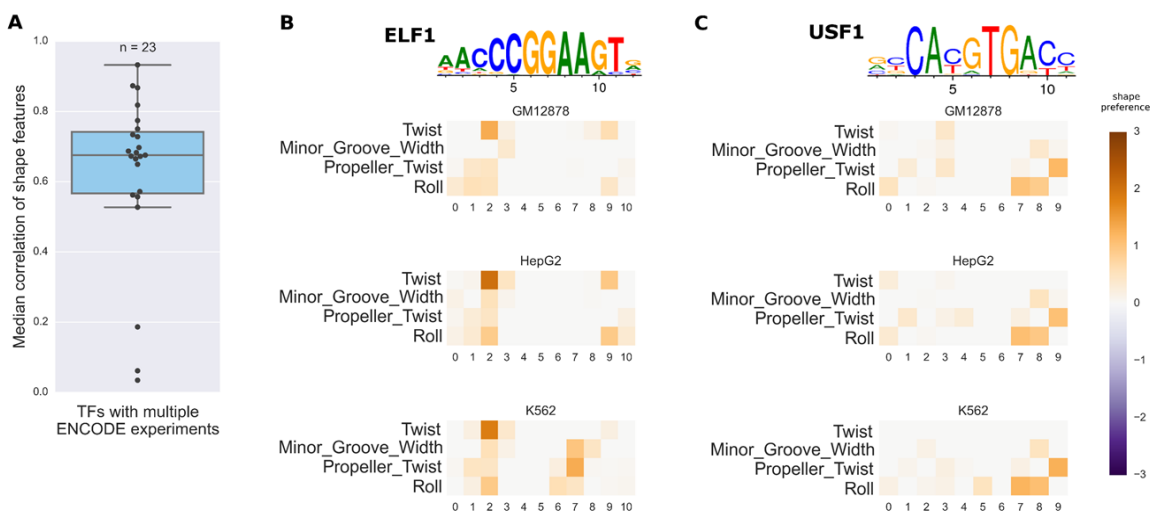

**Figure 3.** Shape model parameters in different conditions. (**A**) Distribution of median correlation coefficients for 23 TFs with multiple ChIP-seq datasets. (**B**) Shape feature preferences for ELF1 in three cell lines. The heatmaps show the preference for each shape feature at each position (the inferred $d^f$ matrix). Only the positive preferences are shown, that is, the preferences towards higher values of shape features. Colors faded to grey mean weaker preference. (**C**) Shape feature

preferences for USF1 in three cell lines. The PWM logos are shown above the predicted shape preference heatmaps.

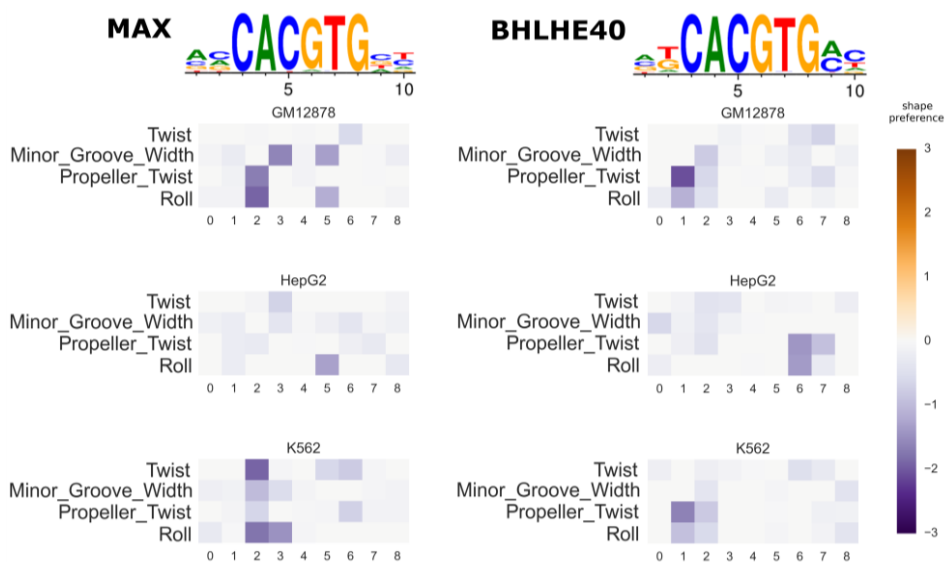

**Figure 4.** Predicted shape preferences of MYC Associated Factor X (MAX) and Basic Helix-Loop-Helix Family Member E40 (BHLHE40) in three cell lines. The heatmaps show the preference for each shape feature at each position (the inferred $d^f$ matrix). Only the negative preferences are shown here, that is the preferences towards lower values of shape features. Colors faded to grey mean weaker preference. The PWM logos are shown above the shape preference heatmaps.

*3.4. Variation of DNA Shape Preferences across Cellular Conditions*

Changing cellular conditions may result in changing gene expression, which is caused by fine-tuning of TF–DNA interaction patterns. This adjustment is tightly controlled and thus may be reflected in the DNA shape feature preferences of affected TFs. Our new model allows identifying such changes by inferring shape preferences from ChIP-Seq data with different conditions. Here, we test the new model on the estradiol (E2)-treated MCF7 breast cancer cell line. It has been observed that the response of treatment includes chromatin reorganization [37], which prompts the investigation of possible DNA shape preference changes in the TFs involved. We used two public ChIP-seq datasets (GEO GSE94023 and GSE72249 [28]) to study this system. In Figure 5A (and Supplementary Figure S3A for negative preferences), the evolution of shape preferences of estrogen receptor α (ERα/ESR1) is shown at different time points after the treatment of MCF7 cells with E2. After E2 stimulation, ERα enters the nucleus and binds to specific sequences in the DNA called estrogen response elements. The shape preferences of ERα remain relatively stable across time. After 320 minutes, the shape preference becomes weak, likely due to reduced signal as there are much fewer strong peaks.

FOXA1 is a pioneer factor, that is it can bind to condensed chromatin and make it accessible to other TFs. As such, it must be precisely targeted to particular DNA sequences. It is known that its binding patterns are affected by the presence of other TFs, in particular ER and the glucocorticoid receptor (GR) [28]. These interactions are critically important for tumorigenesis of breast and prostate cancers [38]. Thus, it is interesting to see whether the change of FOXA1 binding affinity can be observed in the DNA shape preferences or not. In Figure 5B (see Supplementary Figure S3B for positive preferences), the inferred shape preferences of FOXA1 are shown in three conditions of MCF7 cells: untreated, treated with dexamethasone (Dex, a glucocorticoid) and treated with E2. Shape preferences in both treated conditions are similar to each other and slightly different from that in the untreated condition, with additional preferences appearing at dinucleotide 8. This indicates that the modulation of FOXA1 binding by ER and GR involves changes of DNA shape.

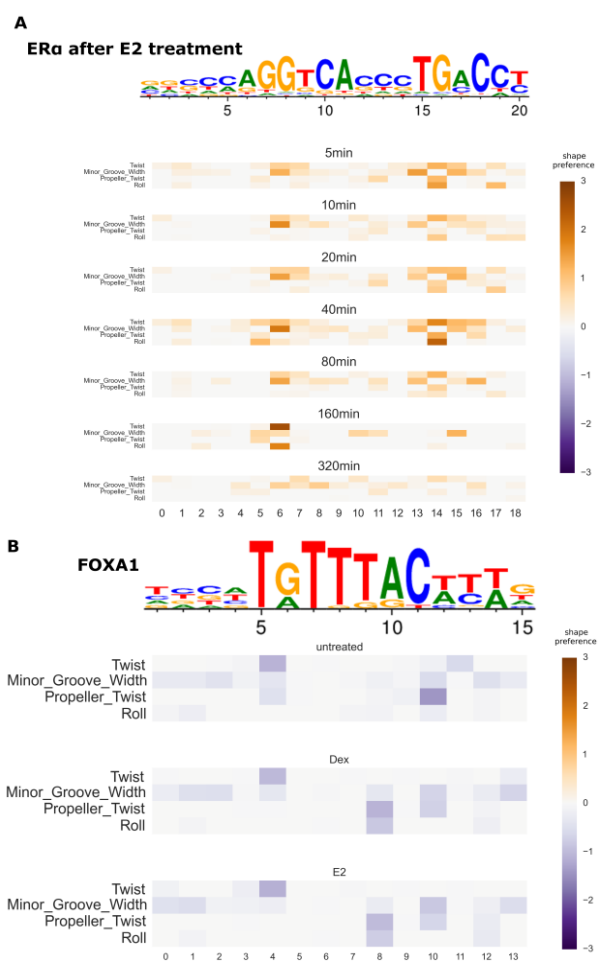

**Figure 5.** Condition-specific DNA shape preferences in protein-DNA interaction. (**A**) Variation of inferred shape preferences of ERα at different time points after the treatment with estradiol (E2) in the MCF7 cell line. (**B**) Variation of inferred shape preferences of FOXA1 after the treatment with either dexamethasone (Dex) or E2 in the MCF7 cell line. The heatmaps show the preference for each shape feature at each position (the inferred $d^f$ matrix). Only the positive preferences are shown in (A), and only the negative preferences are shown in (B). Colors faded to grey mean weaker preference. The PWM logos are shown above the shape preference heatmaps.

*3.5. Allele-Specific Binding Prediction*

We used the BayesPI-BAR algorithm to study whether the prediction of TF binding affinity changes is affected by SNVs. The SNVs used in this study were inferred from the ChIP-seq data, by analyzing the raw reads for the presence of allele-specific binding (ASB) events [29]. There are 36 datasets, one per TF, with SNVs that are marked as either "ASB" (the TF binding is affected by the SNV) or "non-ASB" (no significant effect is observed). The task is to predict whether an SNV causes ASB or not, given the reference and alternate DNA sequence. BayesPI-BAR can solve the problem by predicting the TF binding affinity change between the two sequences with the known TF affinity model. The higher the absolute value of the predicted change, the greater the chance of an ASB event.

We first tested the accuracy of baseline PWM models. BayesPI-BAR uses several alternative PWMs for the same TF simultaneously, in which case the predicted $\delta dbA$ scores for each PWM are averaged. We used several sets of PWMs in this test: (1) 26 PWMs from the JASPAR database, one for every TF is available; (2) 129 PWMs for 28 TFs from the database of 1772 PWMs that was used in the original BayesPI-BAR publication; (3) 112 PWMs for 33 TFs inferred by the BayesPI2 motif discovery program based on ChIP-seq datasets with motif sizes of 10, 15 and 20. The results of these tests are shown in Figure 6. We use the area under the precision-recall curve (AUPRC) as the measure of prediction performance due to class imbalance in the ASB dataset. The median AUPRC of BayesPI-BAR is ≈0.32, which is comparable to the accuracy of other machine learning approaches in [29] (e.g., ≈0.35). However, it is significantly higher than a previous report in [29], where it was ≈0.15. Such a performance discrepancy of BayesPI-BAR in the same datasets may be caused by misusing of principal component analysis (PCA) scores in the earlier work [29], instead of using the mean $\delta dbA$ scores in the current work.

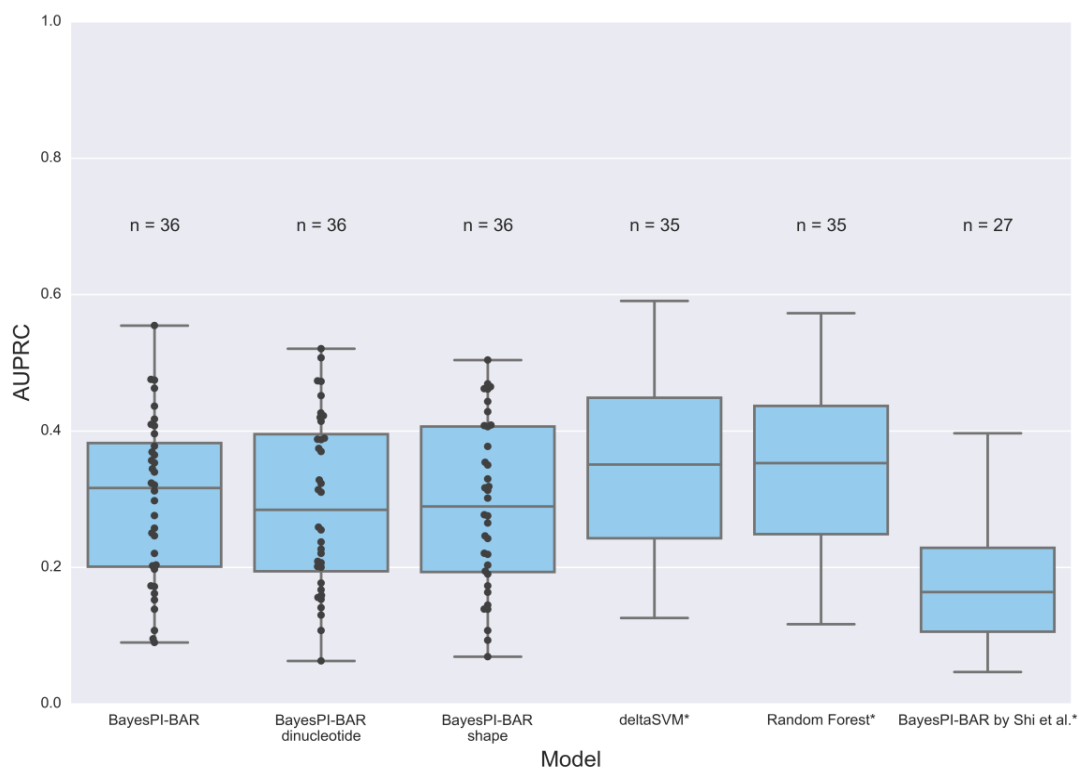

**Figure 6.** Allele-specific binding (ASB) prediction accuracy by various models. Distribution of the area under the precision-recall curve (AUPRC) for predictions of ASB events, for BayesPI-BAR, deltaSVM and a random forest-based sequence model from [29]. The PWM sets used in BayesPI-BAR are: 26 PWMs from the JASPAR database; the relevant PWMs from the set of 1772 human TF PWMs in the original database supplied with BayesPI-BAR; a set of PWMs that could be successfully inferred by the BayesPI motif discovery program using the Encyclopedia of DNA Elements (ENCODE) ChIP-seq data. Data for models marked with a star are reproduced, approximately, from [29].

After confirming that the performance of BayesPI-BAR is satisfactory for the ASB data by considering the independent model only, we compared the performance between the independent and dependent models. For each independent model of a TF, we infer the dependency correction parameters (either the full dinucleotide matrix or the shape feature preferences) using ChIP-seq datasets. The results for the available 36 TFs are summarized in Figure 6 and Supplementary Figure S4. In testing the dependency models, we use the same set of independent models and add the dependency energy terms when they are available. However, neither the full dinucleotide model, nor the DNA shape-restricted model improves ASB prediction accuracy over that by the independent model. For example, there is a big improvement in prediction accuracy in some TFs, but a negative impact on the other TFs. Thus, there is no advantage of using dependent models for ASB event prediction in the current study.

## 4. Discussion

We have developed a new biophysical TF–DNA interaction model that takes into account DNA shape-restricted dinucleotide dependencies. The new model restricts the parameter space of the dinucleotide dependencies to a reduced subspace, which considers only the dinucleotide DNA shape properties. Such an implementation makes the model biophysically interpretable and can be used to investigate the TF–DNA interaction mechanism in various circumstances, by examining the predicted model parameters. It has been previously reported that models including nucleotide dependency parameters can fit datasets (e.g., in protein binding microarray probes, systematic evolution of ligands by exponential enrichment (SELEX) sequences and ChIP-seq peaks) better than the independent models. However, the question of whether the learned dependency features represent the actual TF binding preference or capture a subtle bias of an experimental error remains open. In the DREAM5 (TF-DNA Motif Recognition Challenge) [25], the biophysical models including dinucleotide interactions generally have better performance than the independent ones, when tested on the same type of data such as the in vitro experiment that was used for both training and testing data. On the contrary, the relative ranking of the independent and the dependent models is the opposite when models were trained on in vitro datasets, but tested on in vivo ones. Such a problem is later shown to be solved by a new dinucleotide model (FeatureREDUCE), by using a refined regularization optimization procedure [11]. However, the conclusion of FeatureREDUCE is based only on one TF.

Here, we have used ChIP-seq datasets to fit dinucleotide dependency models by a Bayesian regression procedure, which aims to robustly estimate the model parameters when the input data contain noise. The use of in vivo data for training enables the models to infer the model parameters based on true conditions of TF–DNA binding. For the new shape model, we have compared the inferred shape preferences of a few TFs to that of previously-reported results, which has a reasonable match between the two (Figure 1). The new biophysical model with DNA shape preference features improves the accuracy of ChIP-seq peak prediction, when compared to the baseline independent model (Figure 2). We also looked into the issue of whether DNA shape preferences are largely independent of the cell types or not, by testing the new DNA shape model in ENCODE ChIP-seq data under various cell lines (Figure 3). In some cases, the inferred shape preferences changed between the conditions such as MAX and BHLHE40 TFs (Figure 4). The aforementioned two TFs can bind DNA either individually or as heterodimers with other TFs [39,40], which may explain the variability of the shape preferences under different conditions. Although the core PWM models of MAX and BHLHE40 are very similar, their predicted shape preferences are quite different, which may be used to distinguish the true binding target sites. This is a very interesting observation.

Next, we explored the possibility of the dynamical change of DNA shape preferences in TF–DNA interactions under various conditions, such as the response of MCF7 breast cancer cells to steroid treatment. In the current work, the inferred shape preferences of ER$\alpha$ have little change at different time points after the E2 treatment, except for the last few time points where the number of ChIP-seq peaks is reduced significantly (Figure 5A). Nevertheless, FOXA1 gained new DNA shape dependencies after E2 treatment, and the same additional shape dependencies are also observed after treatment with Dex (Figure 5B). Thus, the target binding sites of FOXA1 have slightly different DNA shape preferences, which indicates TF cofactors are involved in the FOXA1 binding and may influence the DNA geometry differently during the interaction between ER, GR and FOXA1. The new shape model gives us an opportunity to study in detail the intricate TF–DNA interaction patterns and to find a possible role of DNA shapes in different genome regulations.

Finally, we used the BayesPI-BAR framework to assess the power of the DNA shape model in predicting genomic mutation effects. It was tested on an ASB dataset derived from ENCODE ChIP-seq data, in which the BayesPI-BAR with various nucleotide dependency models performed reasonably well (Figure 6). It is worth noting that a clear understanding of the in silico prediction program is needed before applying it on any computational biology problems. For example, based on the same ASB data, a previous study reported a much lower prediction accuracy of BayesPI-BAR than that of the machine learning methods [29], although our reanalysis shows that BayesPI-BAR achieves a similar accuracy as the other methods (Figure 6). This is because the authors of [29] might have used a post-processed TF ranking score (principal component scores, PCA) as the indicator of effect size (TF binding affinity changes). Here, we used the mean $\delta dbA$ (differential binding affinity) as a direct measure of TF binding changes, which correctly represents the TF binding data. The $\delta dbA$ can be reused in the testing set or be compared between different conditions, but this is not the case for PCA scores. Nevertheless, the proposed new shape model did not, on average, improve the prediction accuracy of the mutation effect on TF-DNA binding over that of the independent model, as evidenced by ASB data. This may be due to the limitation of the present model (e.g., BayesPI-BAR does not consider the geometry changes of the DNA molecule when it is wrapped around a nucleosome).

Generally, nucleosome core particles interact with the DNA, bending it nonuniformly and in a sequence-dependent manner, which affects the DNA shape features [35]. The DNA shape features in turn may influence the preferential location of nucleosomes [41]. These interdependencies between the DNA shape and nucleosomes, the dynamic positioning of nucleosomes, the TF-specific interaction with nucleosomes and the lack of true information on nucleosome positions make it difficult to build a biophysical model with the consideration of the nucleosome effects in the TF–DNA interaction. Thus, the nucleosome-related adjustments to the DNA shape are treated as noise in the current model; a future study to overcome this limitation by considering more genomic data and a refined model shall be carried out. Especially, more systematic investigation is needed to access the contribution of DNA shape in regulatory mutation studies.

In conclusion, the new DNA shape enhanced biophysical model enables the investigation of additional aspects of TF–DNA binding. Given the encouraging results in ChIP-seq peak prediction, a further development of the nucleotide dependency models, including the DNA shape preferences model, is a promising direction for future research. Such models, which can substantially improve the prediction accuracy of mutation effects, will lead to a better understanding of mutation-induced genome dysregulation in diseases such as cancer.

**Supplementary Materials:** The following are available online at www.mdpi.com/2073-4425/8/9/233/s1. Figure S1: Fitted shape preferences of MAX and BHLHE40 for different cell lines (positive part). Figure S2: Shape model parameters in different conditions (negative part). Figure S3: Changes in shape preferences. Figure S4: ASB prediction of the full dinucleotide and shape-restricted models compared to the independent model.

**Acknowledgments:** This work was supported by the Norwegian Cancer Society (DNK 2192630-2012-33376, DNK 2192630-2013-33463, and DNK 2192630-2014-33518); the South-Eastern Norway Regional Health Authority (HSØ 2017061); and the Norwegian Research Council NOTUR project (nn4605k).

**Author Contributions:** K.B. carried out data analysis, contributed tools for the study and drafted the manuscript. J.B.W. conceived of and supervised the study. J.B.W. participated in the data analysis, contributing tools and writing the manuscript. All authors read and approved the contents of the final version of the manuscript.

**Conflict of Interests:** The authors declare no conflict of interest.

## References

1. Slattery, M.; Zhou, T.; Yang, L.; Dantas Machado, A.C.; Gordan, R.; Rohs, R. Absence of a simple code: How transcription factors read the genome. *Trends Biochem. Sci.* **2014**, *39*, 381–399.
2. Song, L.; Li, D.; Zeng, X.; Wu, Y.; Guo, L.; Zou, Q. nDNA-Prot: Identification of DNA-binding proteins based on unbalanced classification. *BMC Bioinform.* **2014**, *15*, 298.
3. Wang, J. Quality versus accuracy: Result of a reanalysis of protein-binding microarrays from the DREAM5 challenge by using BayesPI2 including dinucleotide interdependence. *BMC Bioinform.* **2014**, *15*, 289.
4. Wang, J.; Batmanov, K. BayesPI-BAR: A new biophysical model for characterization of regulatory sequence variations. *Nucleic Acids Res.* **2015**, *43*, e147.

5. Abe, N.; Dror, I.; Yang, L.; Slattery, M.; Zhou, T.; Bussemaker, H.J.; Rohs, R.; Mann, R.S. Deconvolving the recognition of DNA shape from sequence. *Cell* **2015**, *161*, 307–318.
6. Lin, C.; Chen, W.; Qiu, C.; Wu, Y.; Krishnan, S.; Zou, Q. LibD3C: Ensemble classifiers with a clustering and dynamic selection strategy. *Neurocomputing* **2014**, *123*, 424–435.
7. Ghandi, M.; Lee, D.; Mohammad-Noori, M.; Beer, M.A. Enhanced regulatory sequence prediction using gapped k-mer features. *PLoS Comput. Biol.* **2014**, *10*, e1003711.
8. Mathelier, A.; Wasserman, W.W. The next generation of transcription factor binding site prediction. *PLoS Comput. Biol.* **2013**, *9*, e1003214.
9. Mathelier, A.; Xin, B.; Chiu, T.P.; Yang, L.; Rohs, R.; Wasserman, W.W. DNA Shape features improve transcription factor binding site predictions in vivo. *Cell Syst.* **2016**, *3*, 278–286.
10. Alipanahi, B.; Delong, A.; Weirauch, M.T.; Frey, B.J. Predicting the sequence specificities of DNA- and RNA-binding proteins by deep learning. *Nat. Biotechnol.* **2015**, *33*, 831–838.
11. Riley, T.R.; Lazarovici, A.; Mann, R.S.; Bussemaker, H.J. Building accurate sequence-to-affinity models from high-throughput in vitro protein-DNA binding data using Feature REDUCE. *eLife* **2015**, *4*, doi:10.7554/eLife.06397.
12. Zhao, Y.; Ruan, S.; Pandey, M.; Stormo, G.D. Improved models for transcription factor binding site identification using nonindependent interactions. *Genetics* **2012**, *191*, 781–790.
13. Wang, J.; Morigen. BayesPI—A new model to study protein-DNA interactions: A case study of condition-specific protein binding parameters for Yeast transcription factors. *BMC Bioinform.* **2009**, *10*, 345.
14. Ramachandran, P.; Palidwor, G.A.; Perkins, T.J. BIDCHIPS: Bias decomposition and removal from ChIP-seq data clarifies true binding signal and its functional correlates. *Epigenetics Chromatin* **2015**, *8*, 33.
15. Orenstein, Y.; Shamir, R. A comparative analysis of transcription factor binding models learned from PBM, HT-SELEX and ChIP data. *Nucleic Acids Res.* **2014**, *42*, e63.
16. Zhao, Y.; Granas, D.; Stormo, G.D. Inferring binding energies from selected binding sites. *PLoS Comput. Biol.* **2009**, *5*, e1000590.
17. Wang, J.; Malecka, A.; Trøenand, G.; Delabie, J. Comprehensive genome-wide transcription factor analysis reveals that a combination of high affinity and low affinity DNA binding is needed for human gene regulation. *BMC Genom.* **2015**, *16* (Suppl. 7), doi:10.1186/1471-2164-16-S7-S12.
18. Batmanov, K.; Wang, W.; Bjørås, M.; Delabie, J.; Wang, J. Integrative whole-genome sequence analysis reveals roles of regulatory mutations in BCL6 and BCL2 in follicular lymphoma. *Sci. Rep.* **2017**, *7*, 7040.
19. Miele, V.; Vaillant, C.; d'Aubenton-Carafa, Y.; Thermes, C.; Grange, T. DNA physical properties determine nucleosome occupancy from yeast to fly. *Nucleic Acids Res.* **2008**, *36*, 3746–3756.
20. Slattery, M.; Riley, T.; Liu, P.; Abe, N.; Gomez-Alcala, P.; Dror, I.; Zhou, T.; Rohs, R.; Honig, B.; Bussemaker, H.J.; et al. Cofactor binding evokes latent differences in DNA binding specificity between Hox proteins. *Cell* **2011**, *147*, 1270–1282.
21. Rohs, R.; West, S.M.; Sosinsky, A.; Liu, P.; Mann, R.S.; Honig, B. The role of DNA shape in protein-DNA recognition. *Nature* **2009**, *461*, 1248–1253.
22. Zhou, T.; Shen, N.; Yang, L.; Abe, N.; Horton, J.; Mann, R.S.; Bussemaker, H.J.; Gordan, R.; Rohs, R. Quantitative modeling of transcription factor binding specificities using DNA shape. *Proc. Natl. Acad. Sci. USA* **2015**, *112*, 4654–4659.
23. Tsai, Z.T.; Shiu, S.H.; Tsai, H.K. Contribution of sequence motif, chromatin state, and DNA structure features to predictive models of transcription factor binding in Yeast. *PLoS Comput. Biol.* **2015**, *11*, e1004418.
24. Yang, J.; Ramsey, S.A. A DNA shape-based regulatory score improves position-weight matrix-based recognition of transcription factor binding sites. *Bioinformatics* **2015**, *31*, 3445–3450.
25. Weirauch, M.T.; Cote, A.; Norel, R.; Annala, M.; Zhao, Y.; Riley, T.R.; Saez-Rodriguez, J.; Cokelaer, T.; Vedenko, A.; Talukder, S.; et al. Evaluation of methods for modeling transcription factor sequence specificity. *Nat. Biotechnol.* **2013**, *31*, 126–134.
26. Friedel, M.; Nikolajewa, S.; Suhnel, J.; Wilhelm, T. DiProDB: A database for dinucleotide properties. *Nucleic Acids Res.* **2009**, *37*, D37–D40.
27. Dunham, I.; Kundaje, A.; Aldred, S.F.; Collins, P.J.; Davis, C.A.; Doyle, F.; Epstein, C.B.; Frietze, S.; Harrow, J.; Kaul, R.; et al. An integrated encyclopedia of DNA elements in the human genome. *Nature* **2012**, *489*, 57–74.
28. Swinstead, E.E.; Miranda, T.B.; Paakinaho, V.; Baek, S.; Goldstein, I.; Hawkins, M.; Karpova, T.S.; Ball, D.; Mazza, D.; Lavis, L.D.; et al. Steroid receptors reprogram FoxA1 occupancy through dynamic chromatin transitions. *Cell* **2016**, *165*, 593–605.

29. Shi, W.; Fornes, O.; Mathelier, A.; Wasserman, W.W. Evaluating the impact of single nucleotide variants on transcription factor binding. *Nucleic Acids Res.* **2016**, *44*, 10106–10116.
30. Zhou, T.; Yang, L.; Lu, Y.; Dror, I.; Dantas Machado, A.C.; Ghane, T.; Di Felice, R.; Rohs, R. DNAshape: A method for the high-throughput prediction of DNA structural features on a genomic scale. *Nucleic Acids Res.* **2013**, *41*, W56–W62.
31. Wang, J. The effect of prior assumptions over the weights in BayesPI with application to study protein–DNA interactions from ChIP-based high-throughput data. *BMC Bioinform.* **2010**, *11*, 412.
32. Mackay, D. Bayesian methods for adaptive models. Ph.D. Thesis, California Institute of Technology, Pasadena, CA, USA, 1991.
33. Bewley, C.A.; Gronenborn, A.M.; Clore, G.M. Minor groove-binding architectural proteins: Structure, function, and DNA recognition. *Annu. Rev. Biophys. Biomol. Struct.* **1998**, *27*, 105–131.
34. Lazarovici, A.; Zhou, T.; Shafer, A.; Dantas Machado, A.C.; Riley, T.R.; Sandstrom, R.; Sabo, P.J.; Lu, Y.; Rohs, R.; Stamatoyannopoulos, J.A.; et al. Probing DNA shape and methylation state on a genomic scale with DNase I. *Proc. Natl. Acad. Sci. USA* **2013**, *110*, 6376–6381.
35. Segal, E.; Fondufe-Mittendorf, Y.; Chen, L.; Thastrom, A.; Field, Y.; Moore, I.K.; Wang, J.P.; Widom, J. A genomic code for nucleosome positioning. *Nature* **2006**, *442*, 772–778.
36. Mathelier, A.; Fornes, O.; Arenillas, D.J.; Chen, C.Y.; Denay, G.; Lee, J.; Shi, W.; Shyr, C.; Tan, G.; Worsley-Hunt, R.; et al. JASPAR 2016: A major expansion and update of the open-access database of transcription factor binding profiles. *Nucleic Acids Res.* **2016**, *44*, D110–D115.
37. Wang, J.; Lan, X.; Hsu, P.Y.; Hsu, H.K.; Huang, K.; Parvin, J.; Huang, T.H.; Jin, V.X. Genome-wide analysis uncovers high frequency, strong differential chromosomal interactions and their associated epigenetic patterns in E2-mediated gene regulation. *BMC Genom.* **2013**, *14*, 70.
38. Jozwik, K.M.; Carroll, J.S. Pioneer factors in hormone-dependent cancers. *Nat. Rev. Cancer* **2012**, *12*, 381–385.
39. Naud, J.F.; McDuff, F.O.; Sauve, S.; Montagne, M.; Webb, B.A.; Smith, S.P.; Chabot, B.; Lavigne, P. Structural and thermodynamical characterization of the complete p21 gene product of Max. *Biochemistry* **2005**, *44*, 12746–12758.
40. Sato, F.; Kawamoto, T.; Fujimoto, K.; Noshiro, M.; Honda, K.K.; Honma, S.; Honma, K.; Kato, Y. Functional analysis of the basic helix-loop-helix transcription factor DEC1 in circadian regulation. Interaction with BMAL1. *Eur. J. Biochem.* **2004**, *271*, 4409–4419.
41. Bolshoy, A. CC dinucleotides contribute to the bending of DNA in chromatin. *Nat. Struct. Biol.* **1995**, *2*, 446–448.